\documentclass[paper=a4, fontsize=11pt]{article}
\usepackage{iftex,amsmath}
\iftutex
\usepackage{unicode-math}
\else
\usepackage{amssymb}
\def\z"{}
\def\UnicodeMathSymbol#1#2#3#4{%
 \ifnum#1>"A0
   \DeclareUnicodeCharacter{\z#1}{#2}%
  \fi}
\input{unicode-math-table}

\fi
\usepackage{amsthm}
\usepackage{enumitem}
\usepackage{cancel}
\usepackage{appendix}
\usepackage{graphicx}
\usepackage{geometry}
\usepackage{gensymb}
\usepackage{url}
\usepackage[colorlinks=true,linkcolor=blue]{hyperref}
\usepackage{sectsty}
\allsectionsfont{ \normalfont\scshape \textbf}

\newlist{steps}{enumerate}{1}
\setlist[steps, 1]{label = Step \arabic*:}
\numberwithin{equation}{section}
\title{New plasmon-like mode in PdTe$_{2}$: Raman scattering and memory function study }
\author{Bharathiganesh Devanarayanan $^{a,b}$$\,^{*}$, Sahil Rathi$^{a,b}$, Jalaja Pandya$^{a}$,\\ Sonika$^{c}$, C.S.Yadav$^{c}$,  Navinder Singh$^{a}$, Satyendra Nath Gupta$^{a,**}$\\
        \small $^{a}$ Physical Research Laboratory, \\\small Navrangpura Ahmedabad, India - 380009 \\
        \small $^{b}$Indian Institute of Technology, Gandhinagar, Palaj, Gujarat, India - 382355 \\
        \small $^{c}$ School of Physical Sciences, Indian Institute of Technology Mandi,\\\small Kamand, Mandi, Himachal Pradesh, India -175075 \\\\
        \small Corresponding authors:\,$^{*}$\tt{dbharathiganesh@gmail.com},\,$^{**}$\tt{satyendra@prl.res.in}
}
\date{} 

\begin{document}
\maketitle
\begin{abstract}
PdTe$_2$ is a type II Dirac semimetal that has garnered significant attention due to its intriguing electronic and topological properties. Here, we report temperature dependent Raman scattering study of PdTe$_2$ in the temperature range from 10 K to 300 K.  Our study reveals emergence of a new unreported peak below 100 K, centered around 250 cm$^{-1}$.   We argue that the new mode is not a phonon mode because the Raman spectra calculated using Density Functional Theory shows only two intense peaks at 85 $ cm^{-1}$ and 128 $cm^{-1}$. To ascertain the origin of this new peak, we constructed a microscopic model of electrons coupling to a single plasmon mode at 250 $cm^{-1}$ and using the memory function formalism, we obtained that the Raman relaxation rate is linear in frequency. We also performed phenomenological analysis of the Raman response from the experimental data and computed frequency dependent Raman relaxation rate, which is also found to exhibit a linear dependence on frequency. With the congruence of our theoretical and phenomenological results we could ascertain that the new mode observed at low temperatures is indeed a plasmon-like mode.  Further, phonon frequencies and line widths of the  two phonon modes exhibit anomalous behavior above 100 K.     
 
\noindent   \end{abstract}

\section{Introduction}\label{sec1}


Dirac semimetals are fascinating materials with Dirac points where the conduction band and valence band meet each other and is a four fold degenerate band crossing. The electronic dispersion near this Dirac point is linear and form the ``Dirac cones'' above and below the Dirac point. If the Dirac cone is tilted and anisotropic in the momentum space, the material is called a type II Dirac semimetal.  One such example is  Palladium di Telluride (PdTe$_2$), which is a Transition Metal Dichalcogenide (TMDC) that has recently garnered significant scientific attention due to various interesting physical properties it exhibits\cite{Pdte2_int_1,csy2}. It shows superconductivity with a T$_{c}$ of 1.7 K\cite{csy3,csy4} and is reported to be a type II Dirac semimetal with tilted Dirac cones\cite{Pdte2_int_1,Pdte2_intro_1} and with nontrivial topological surface states\cite{TMDC_intro_3,TMDC_intro_10,TMDC_intro_12}.\\

Although, there have been extensive studies of electronic properties of PdTe$_2$, the properties related to lattice vibrations, electron-phonon coupling etc. are still not much explored.  Raman spectroscopy\cite{RAMAN1928,Raman_intro_1} is a powerful non-destructive analytical tool to provide detailed insights into the lattice vibrations\cite{Raman_intro}, electron-phonon interactions\cite{Raman_intro_2,Raman_intro_3}, plasmons\cite{plasmon,plasmon2} and exotic quantum excitations of materials such as topological insulators\cite{Raman_intro_5,Raman_intro_4,lemmen3}, quantum spin liquids\cite{Raman_intro_6}, high-temperature superconductors\cite{Raman_intro_3},2D materials\cite{Raman_intro_7,Raman_intro_8,lemmen2} etc. Raman spectroscopy performed at low temperature provides additional information in the spectra that might be masked at room temperature\cite{Taylor:51,lemmen1,cd3as2,Huang2016}.\\

Raman scattering experiments performed at low temperatures on another Dirac material (Cd$_3$As$_2$) report the emergence of a new Raman mode around 250 $cm^{-1}$ below a characteristic temperature of 100 K\cite{cd3as2}. However, a reasonable understanding of the origin of this new Raman mode seems to be missing in the literature. It should also be noted that anomalous behavior around the temperature of 100 K in these systems has been highlighted in other experiments too like in the case of thermal transport\cite{csyadav1}.  \\ 


Here, we report a detailed study of PdTe$_2$ using temperature dependent Raman spectroscopy at various temperatures ranging from as low as 10 K to the room temperature. We note the emergence of a new peak in the Raman spectra of PdTe$_2$ at lower temperatures ($ \lesssim  100 K$) and analyze the behavior of the peak position and width as a function of temperature. This new peak arises at around 250 cm$^{-1}$ and is broader than the other two phonon peaks centered around 125 cm$^{-1}$ and 165 cm$^{-1}$. The other two Raman modes also exhibit anomalous behavior above 100 K in their phonon frequencies and line widths.\\

Our Density Functional Theory (DFT) calculations reveal that there are only two phonon modes in PdTe$_2$ implying that the new peak possibly corresponds to a plasmon-like mode emerging below 100 K\cite{cd3as2}. To ascertain this, we perform both theoretical calculations with a simple microscopic model of electrons coupled to a single plasmon mode\cite{mmformalism_2} using the memory function formalism and phenomenological analysis\cite{mmformalim_1} to obtain the Raman relaxation rate and the mass enhancement factor from the experimentally obtained Raman spectra. We observed in both the theoretical and phenomenological analysis that the Raman relaxation rate has a linear behavior w.r.t. frequency for temperatures below 100 K (with the presence of the third peak) which clearly indicates a Plasmon-like mode in the system as evident from the results of the simple microscopic model. \\

The article is organized as follows. In Sec.\ref{sec2}, we discuss the details of the experiments performed to obtain the results reported in this article. We follow this by presenting the experimental, computational, theoretical and phenomenological results in Sec.\ref{sec4}. We then discuss in detail the implications of our results in Sec.\ref{sec:dis}. We conclude the article in Sec.\ref{sec5}. 

\section{Experimental details}\label{sec2}


Single-crystals of PdTe$_2$ were synthesized using the solid-state reaction method. First, the polycrystalline samples were prepared, taking the granules of Pd ($99.95\%$), and Te ($99.999\%$) in the correct stoichiometric ratio. The reactants were then grounded, pelletized and were sealed in an evacuated quartz tube with pressure lower than $10^{-4}\, mbar$. In the second step, the tube that contained the sample was heated to $850\degree C$ for $48 h$ and subsequently cooled to $550\degree C$ at a very slow cooling rate of $2.5 \degree C \,per\, hour$. It was then kept at $550\degree C$ for another $48 h$ and was naturally cooled down to room temperature. The obtained single crystals were of dimensions $\sim 2 × 5 \,mm^{2}$, which were oriented in the ab-planes. The phase purity and composition of the compounds were checked using x-ray diffraction and SEM-EDAX; and these results are reported elsewhere\cite{csyadav1}.\\

Raman measurements were carried out at various temperatures in backscattering configuration with micro-optical system equipped with Renishaw inVia Raman microscope coupled with a  Peltier-cooled charge controlled detector and 532-nm diode laser. Single crystals of PdTe$_2$ were placed inside an optical cryostat coupled with close cycle Helium compressor system. Temperature variation was done using a controlled heater (Lakeshore) with temperature accuracy of 0.1 K.




\section{Results}\label{sec4}

\subsection{Experimental Results}
\begin{figure}
    \centering
    \includegraphics[width=0.8\linewidth]{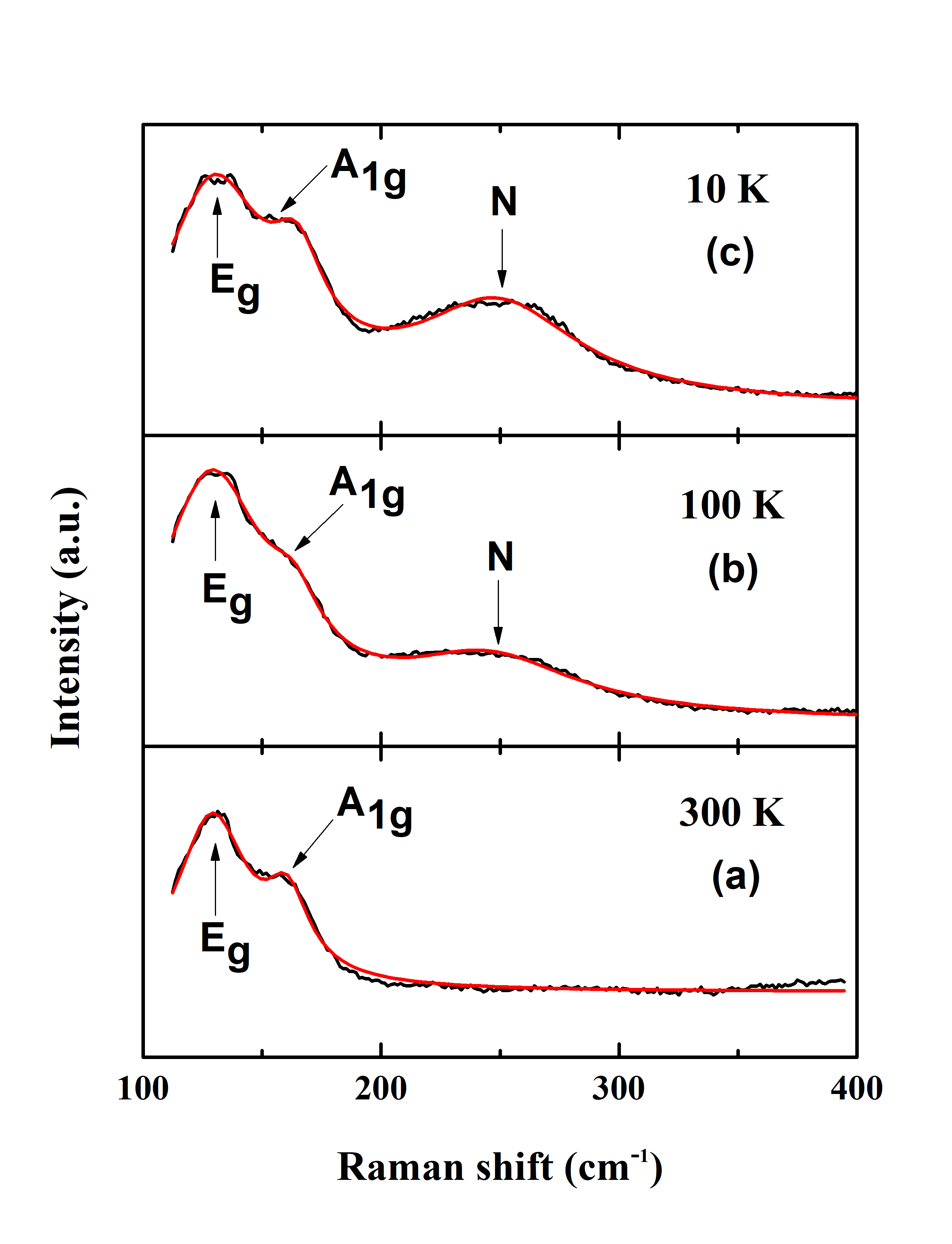}
    \caption{Measured intensity (a.u.) (black) as a function of Raman shift ($cm^{-1}$) and corresponding Lorentzian fit (red) for three different values of temperature (a) 300 K, (b) 100 K and (c) 10 K. The two peaks E$_{g}$ and A$_{1g}$ have already been reported in the literature. However at lower temperatures there is another new peak emerging which we have denoted as ``N''.   }
    \label{int_rs}
\end{figure}

  Fig.\ref{int_rs} shows the Raman spectra of PdTe$_2$ at 300K (Fig. \ref{int_rs} a). We have observed two Raman modes, which were assigned to A$_{1g}$ and  E$_g$ modes based on our DFT calculation as discussed in the DFT section below.  The temperature dependent evolution of Raman spectra is shown in Fig.\ref{int_rs}. 
It is important to notice a profound third peak centered around $250\, cm^{-1}$ at 10K and this new peak is visibly broader than the two existing Raman modes. For a characteristic temperature value (100 K), the position and shape of the first two peaks remains relatively unchanged but the intensity of the third peak has gone down appreciably. For higher temperatures (300 K), the first two peaks persist without much changes but the third peak has disappeared completely.\\

\begin{figure}
    \centering
    \includegraphics[width=0.8\linewidth]{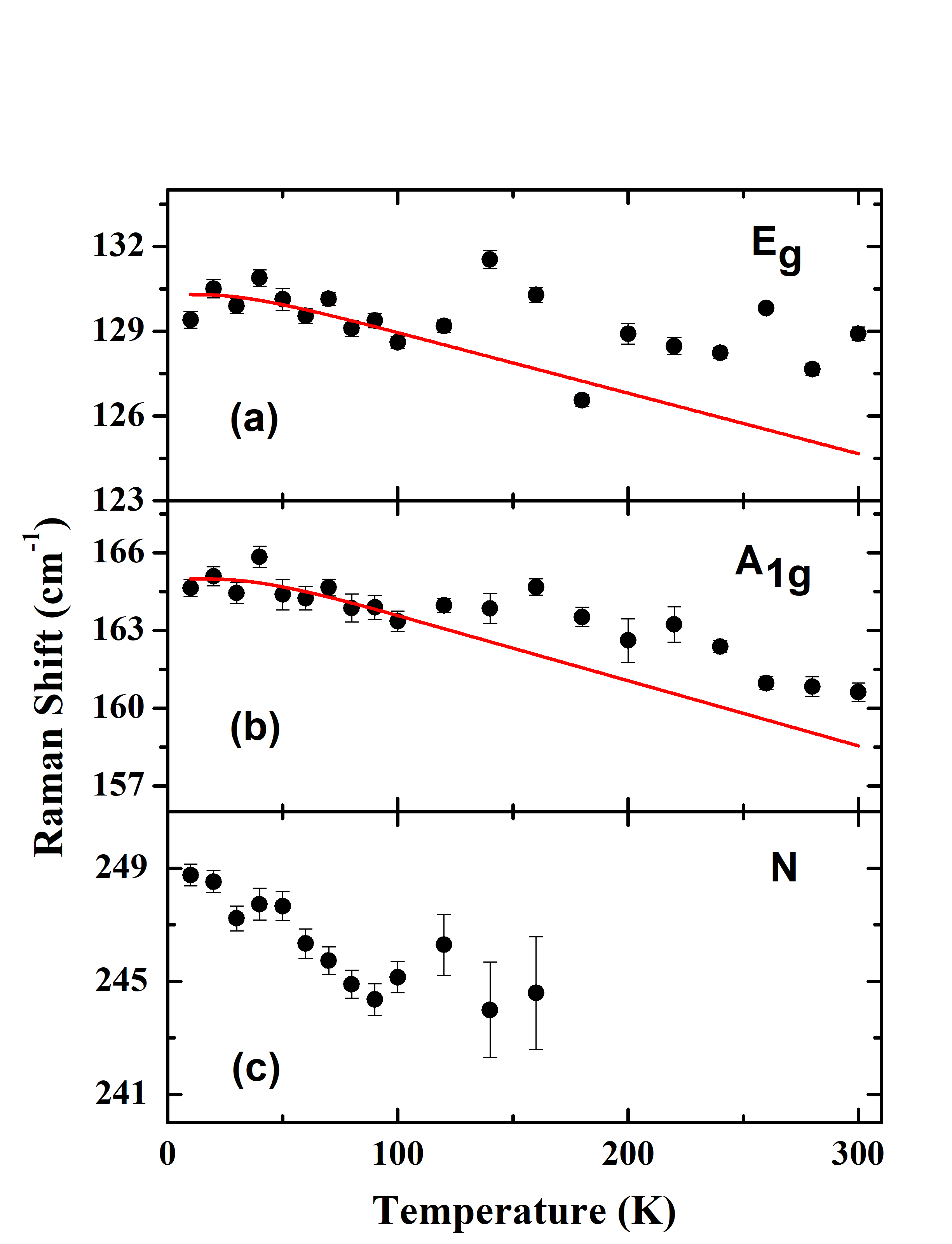}
    \caption{The center of peak position obtained from Lorentzian fit of data in Fig.\ref{int_rs} as a function of temperature (black dots) for (a) E$_{g}$ peak, (b) A$_{1g}$ peak and (c) N peak. The redline is the anharmonicity fit.}
    \label{pp_T}
\end{figure}

\begin{figure}
    \centering
    \includegraphics[width=0.8\linewidth]{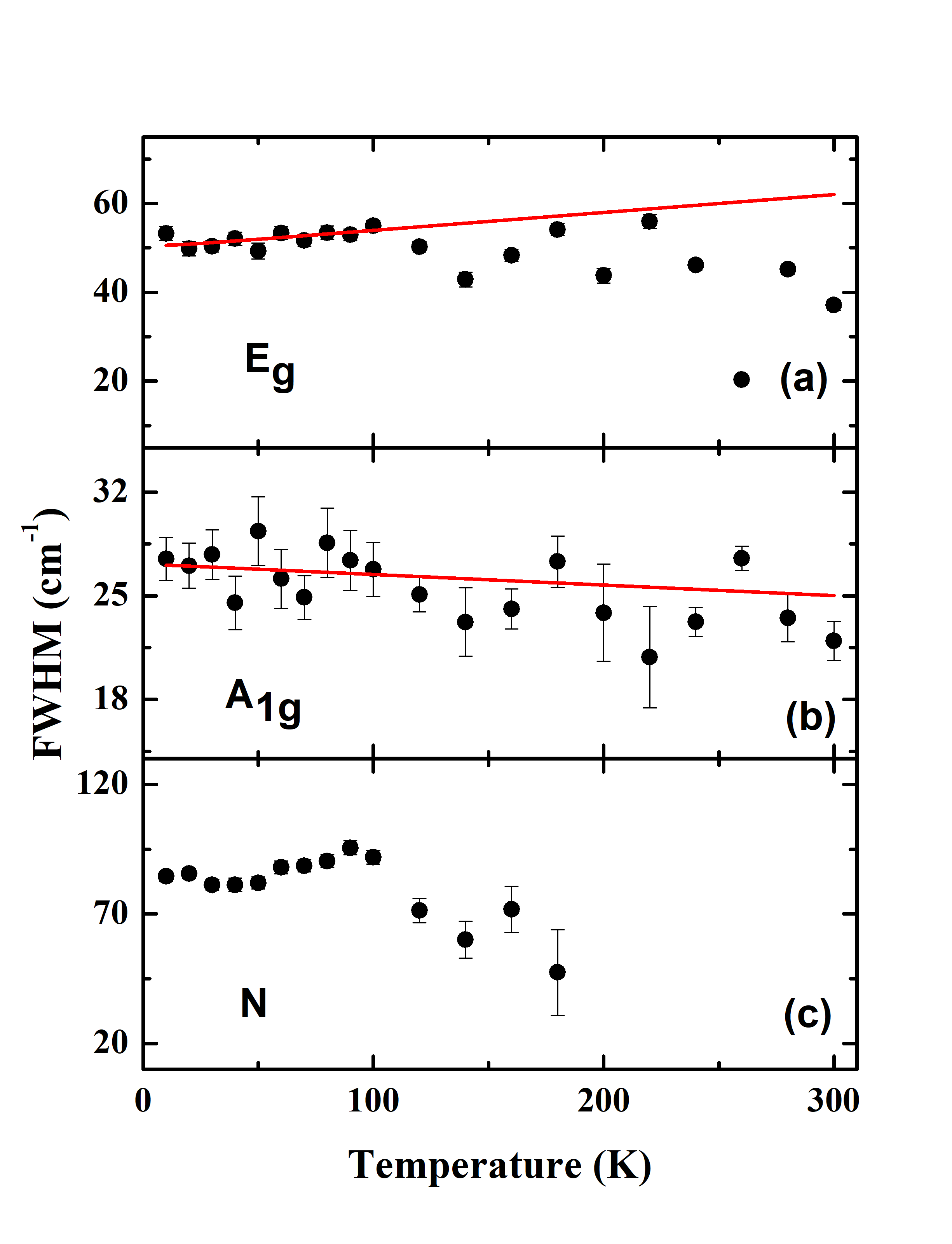}
    \caption{The Full Width at Half Maxima (FWHM) of the peaks obtained from Lorentzian fit of data in Fig.\ref{int_rs} as a  function of temperature (black dots) for (a) E$_{g}$ peak, (b) A$_{1g}$ peak and (c) N peak. The redline is the anharmonicity fit.}
    \label{pw_T}
\end{figure}

This clearly indicates the appearance of a new Raman mode for PdTe$_2$ at lower temperatures which has not been reported in the literature before. To further analyze this new mode and compare it with the other two modes, we fitted the Raman spectrum with Lorentzian line shapes in order to get the  phonon frequencies,  full width half maximum (FWHM) and mode intensities. The temperature dependence of  peak positions and peak widths of the three peaks are shown in Fig.\ref{pp_T} and Fig.\ref{pw_T} respectively. It is clear that the peak position of this new mode shifts towards lower wavenumber as temperature is increased. FWHM of this new mode exhibits slight increase as the temperature is increased. The frequencies and FWHM of the two phonon modes soften and broadens respectively as temperature is increased. We modeled the temperature dependence of these modes with the  simple cubic anharmonicity model,  where the phonon decays into two phonons of equal energy, giving a temperature dependence of phonon frequency and FWMH by 

\begin{equation}
	\omega^{cubic}(T) = \omega(0)+C[1+2n(\omega(0)/2)]
\end{equation}

\begin{equation}
	\Gamma^{cubic}(T) = \Gamma (0)+ D[1+2n(\omega(0)/2)]
\end{equation}

 where $\omega(0)$ is the frequency at absolute zero and $n(\omega) = 1/(exp(\hbar\omega/k_BT)-1)$ is the Bose-Einstein mean occupation number. $\hbar$ and $k_B$ are Plank's and Boltzman's constant respectively.  $\Gamma (0)$ is the  line width at absolute zero. Here, C and D are assumed to be constant depending on the three phonon interaction vertex. The coefficient C is usually negative, i.e. phonon frequency decreases with increasing temperature. The coefficient D is usually  positive, i.e. phonon line width increases with increasing temperature. The model fits  the frequency and FWHM data very well till 100K and show slight deviation above 100K.

\subsection{DFT Calculations}\label{DFT}
Our first principles calculations based on the Density Functional Theory (DFT) were performed using the projected augmented-wave method implemented within the QUANTUM ESPRESSO package \cite{QE-2009}. The simulations were performed within the Generalized Gradient Approximation (GGA) incorportating the spin orbit coupling (SOC). The structure parameters of PdTe$_2$ were obtained from the materials project [https://next-gen.materialsproject.org]. The plane wave kinetic energy cut off was taken to be 70 Ry. The k mesh of 8x8x8 was used to sample the Brillouin zone. Phonon and Raman spectra are calculated by density functional perturbation theory (DFPT). 
\\

\begin{figure}
    \centering
    \includegraphics[width=0.8\linewidth]{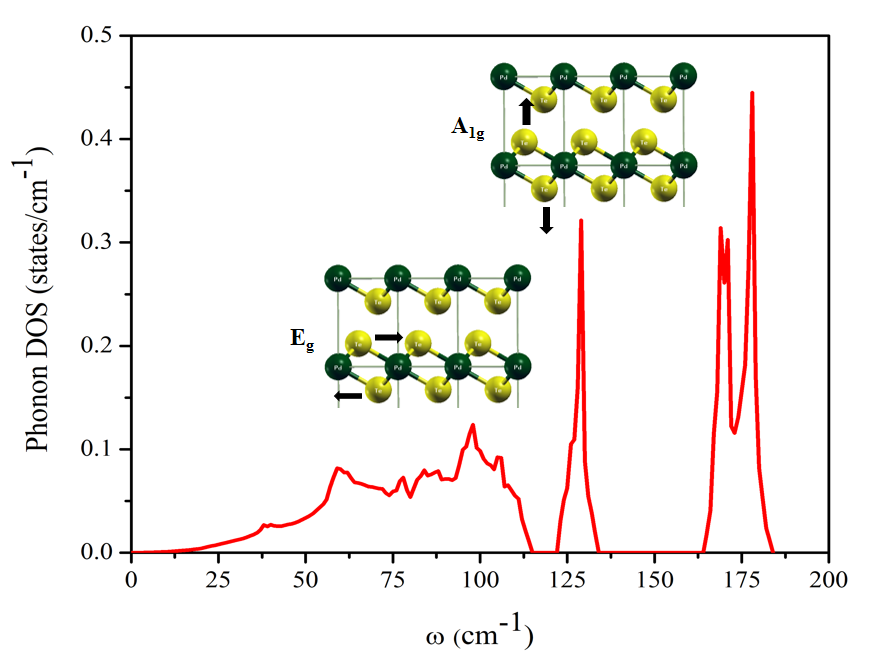}
    \caption{PhDOS of PdTe$_2$ with the Raman modes eigenvectors.}
    \label{Ph-dos}
\end{figure}

To analyze the phonon property of PdTe$_2$, we first optimized the crystal structure. As discussed in Sec. \ref{sec1} it is a layered octahedron structure with space group P-3m1 (No. 164). One Pd atom has 6 Te atoms as its nearest neighbours forming an octahedron (The structure is shown in Fig. \ref{Ph-dos}). The c/a ratio is $\sim 1.27$ which is less than that of a typical hcp structure.
    
\

The dynamic stability of the structure is confirmed from the phonon calculations which give the vibrational spectra with absence of imaginary frequencies. There are nine vibration modes, of which three are acoustic and six are optical. The phonon calculations performed at $\Gamma$ reveal two Raman modes at 85 $cm^{-1}$ and 128 $cm^{-1}$ corresponding to E$_g$ and A$_{1g}$ modes respectively. Based on this calculations, we assigned the two observed Raman modes in the experiments as  E$_g$ and A$_{1g}$ modes. Out of the nine vibrational modes, two are E$_g$ modes and one is A$_{1g}$ mode. The same two Raman modes are also observed in the case of 2D-PdTe$_2$ (monolayer) from DFT simulations \cite{sharma2020investigating}. The E${_g}$ mode is degenerate with in-plane vibrations, whereas the A$_{1g}$ mode is nondegenerate with out-of-plane vibrations. The eigenvectors corresponding to both Raman modes are displayed in Fig.\ref{Ph-dos} along with the Phonon Density of State (PhDOS) of PdTe$_2$. 

\subsection{Theoretical model: Memory function formalism}

Based on our analysis of the experimental data of the Raman spectra at various temperatures, it is clear that a prominent third peak in the spectra develops below the characteristic temperature of 100 K. Our DFT calculations discussed in Sec.\ref{DFT}, indicates that in PdTe$_2$ there are only two Raman modes (E$_g$ and A$_{1g}$) that can be assigned to phonons. This clearly hints that the third mode is not a phonon mode and since the system is non-magnetic it cannot be a magnon mode either. A survey of literature of Raman studies on similar systems revealed appearance of a similar peak centered around $250\,cm^{-1}$ and is termed as ``plasmon-like"\cite{cd3as2}. However there are no clear microscopic or phenomenological model available in the literature to back this claim. We would like to continue with this assumption and compute the Raman relaxation rate from the imaginary part of the memory function. We will then compare it with the scattering rate directly obtained from the experiment (Raman spectra). Reasonable agreement between the theoretically calculated and phenomenologically obtained Raman relaxation rates will vindicate the validity of our assumption. \\

In this section we will discuss a microscopic model based on the memory function formalism\cite{mmformalism_2,navindersir} with the presence of a single plasmon mode in the system. We consider only one plasmon mode as we are informed apriori from the experiments that there is only one plasmon mode at around 250 $cm^{-1}$ wavenumber. The Hamiltonian of the model is given as follows:
 \begin{eqnarray}
            H &=& H_{0} + H_{el-plas},\\
            H_{0} &=& \underset{K}{\Sigma} \,\varepsilon_{k} \,c_{k}^{\dagger}c_{k},\\
            H_{el-plas} &=& \underset{k,k'}{\Sigma}\,\lambda(c_{k}^{\dagger}c_{k'}\,b_{k-k'}+H.C.), 
 \end{eqnarray}
 where, $H_{0}$ is the free electron part, $c^{\dagger}_{k}(c_{k})$ are the electronic creation (annihilation) operators, $b^{\dagger}_{k}(b_{k})$ are the plasmonic creation (annihilation) operators and $\lambda$ is the electron-plasmon coupling strength.

In the memory function formalism, the dynamical response of a system is expressed in terms of a relaxation or memory function $M(z)$. For the case of Raman response, the complex Raman response function $\chi(z)$ as a function of the memory function $M(z)$ is given as\cite{mmformalism_2,mmformalim_1,navindersir}:
\begin{eqnarray}
    \chi(z) = \frac{M(z)}{z + M(z)}.
\end{eqnarray}
Due to the asymptotic behavior and the symmetry properties ($M(z) = -M(-z)$ and $M^*(z) = M(z^*)$) of the memory function, it can be represented by a spectral function $M''(\omega)$ as:
\begin{eqnarray}
    M(z) = \frac{1}{\pi} \int_{-\infty}^{+\infty} d\xi \,\frac{M''(\xi)}{\xi-z},
\end{eqnarray}
where $M''(\omega)$ is the analytic continuation of $M(z)$ to the real axis, implying:
\begin{eqnarray}
    M(\omega \,\pm\, i\delta) = M'(\omega) \,\pm\, iM''(\omega.) 
\end{eqnarray}
For the case of real frequencies $\omega$, from the symmetry properties of the memory function ($M(z)$), we see that $M'(\omega)$ and $M''(\omega)$ are odd and even functions, respectively. With this, the generally followed notation for $M(\omega)$ is:
\begin{eqnarray}
    M(\omega) = \omega\,\lambda(\omega)\, +\,i\,\Gamma(\omega), 
\end{eqnarray}
where, both the functions $\lambda(\omega)$ and $\Gamma(\omega)$ are even. $\lambda(\omega)$ describes the frequency dependent mass enhancement, with $1 + \lambda(\omega) = m^*/m_b$, here $m_b$ is the band mass. $\Gamma(\omega)$ can be interpreted as the inverse of a frequency dependent Raman relaxation time (or Raman relaxation rate).\\

With this introduction, it is clear that the imaginary part of the memory function gives the Raman relaxation rate. The expression for calculating the imaginary part of the memory function that has been derived by us (analogous to the case of electrons coupling to phonons in a system) is given as follows\cite{mmformalism_2,mmformalim_1,navindersir}:
 \begin{eqnarray}
        M''(\omega) = \frac{2\pi}{3}\frac{\lambda^2}{ne^2m^*}\,\underset{k,k'}{\Sigma}\,|k-k'|^{2}\,(1-f_{k})f_{k'}\,n_{k-k'}\nonumber\\\left(\frac{e^{\beta \omega }-1}{\omega}\,\delta\left(\varepsilon_k-\varepsilon_{k'}-\omega_{k-k'}+\omega\right) - (terms \,with\,\omega\, \rightarrow -\omega)\right).
 \end{eqnarray}

Imposing that there is only one plasmon mode along with energy and momentum conservation and with some further simplification, we get:

 \begin{eqnarray}\label{mm_eqn}
        M''(\omega) &=& \frac{e^{\beta \omega}-1}{\omega} \int_{0}^{\infty} d\varepsilon [\sqrt{\varepsilon}\sqrt{\varepsilon-\omega_{0}+\omega}\,\theta(\varepsilon-\omega_{0}+\omega)\nonumber\\&&\left(1-\frac{1}{e^{\beta(\varepsilon-\mu)}+1}\right)\left(\frac{1}{e^{\beta(\varepsilon-\omega_{0}+\omega-\mu)}+1}\right) + (terms\, with\,\omega\, \rightarrow-\omega)].
\end{eqnarray}

\begin{figure}
    \centering    \includegraphics[width=0.65\linewidth]{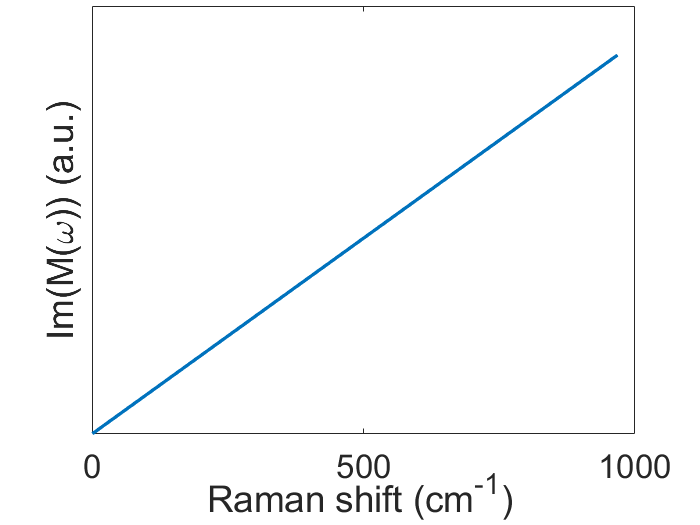}
    \caption{Imaginary part of the memory function as a function of wavenumber indicating a linear relation between both.}
    \label{plas_mm}
\end{figure}

Numerically integrating the integrals in Eq. \ref{mm_eqn}, we can obtain the imaginary part of the memory function as a function of the wavenumber. With the obtained imaginary part of the memory function, plotting it at the region of interest, i.e. between 100-1000 $cm^{-1}$ wavenumber at a temperature below 100 K ($\sim 10$ K), we get a linear relation between the imaginary part of the memory function and the wavenumber as indicated in Fig. \ref{plas_mm}.\\

It is important to recall here that the expression for the memory function is: $M(\omega) = \omega \lambda(\omega) \,+\, i\,\Gamma(\omega)$
(where, $\Gamma(\omega)$ is the Raman relaxation rate and $\lambda(\omega)$ is the mass enhancement factor). From our microscopic model we have found that the Raman relaxation rate has a linear dependence on the wavenumber in the region of our interest where our experimental results were obtained. If we can phenomenologically calculate the relaxation rate from the experimental data and find it to have a linear dependence with the wavenumber, we can decisively conclude that the third peak is plasmon-like in origin. For this purpose, we will discuss the phenomenological analysis in the next section.

\subsection{Phenomenological Analysis}

In this section we will discuss the phenomenological analysis of the experimental data. We will follow the formalism introduced by M.Opel et al. in \cite{mmformalim_1} to analyze the Electronic Raman Spectra (ERS) data, which is different from the original formulation of G$\Ddot{o}$tze and W$\Ddot{o}$lfle\cite{mmformalism_2}.  We will present the calculated values of the relaxation rate ($\Gamma(\omega)$) and the mass enhancement factor ($1 + \lambda(\omega) = m^{*}/m_b$), where $m_b$ is the band mass. \\

In the memory function formalism, the correlation function $I(\omega,T)$ is related to the measured Raman spectrum which is defined as follows:

\begin{equation}
    I(\omega,T) = \frac{\dot{N}(\omega,T)}{\omega\{1\,+ n_{B}(\omega,T)\}},
\end{equation}

where, $\dot{N}(\omega,T)$ is the experimentally measured Raman spectra and $n_{B}(\omega,T)$ is the Bose-Einstein function, defined as:
\begin{equation}
    n_{b}(\omega,T) = 1/(exp((\hbar \omega/k_B T) -1).
\end{equation}

After utilizing the experimental data and obtaining $I(\omega,T)$, one can obtain the Raman relaxation rate ($\Gamma(\omega)$) and the mass-enhancement factor ($1 + \lambda(\omega)$) from the following expressions:

\begin{eqnarray}
    \Gamma(\omega) = R\frac{I(\omega)}{[I(\omega)]^{2} + [\omega K(\omega)]^2},\\
    1 +\lambda(\omega) = R\frac{K(\omega)}{[I(\omega)]^{2} + [\omega K(\omega)]^2},
\end{eqnarray}

where $\omega K(\omega,T)$ is the Kramer-Kronig transform of $I(\omega,T)$; hence the expression for $K(\omega,T)$ is as follows:

\begin{equation}
    K(\omega) = -\frac{2}{\pi}\mathcal{P}\int_{0}^{\omega_c} d\xi \frac{I(\xi)}{\xi^{2} - \omega^{2}}, 
\end{equation}

where $\mathcal{P}$ indicates the principle value integral and the normalizing factor $R$ is fixed by the following sum rule:

\begin{equation}
    R = \frac{2}{\pi}\int_{0}^{\omega_c}d\omega I(\omega).
\end{equation}
In the above integrals $\omega_c$ is the upper cut-off frequency for performing the numerical integration.\\

\begin{figure}
    \centering
    \includegraphics[width=0.65\linewidth]{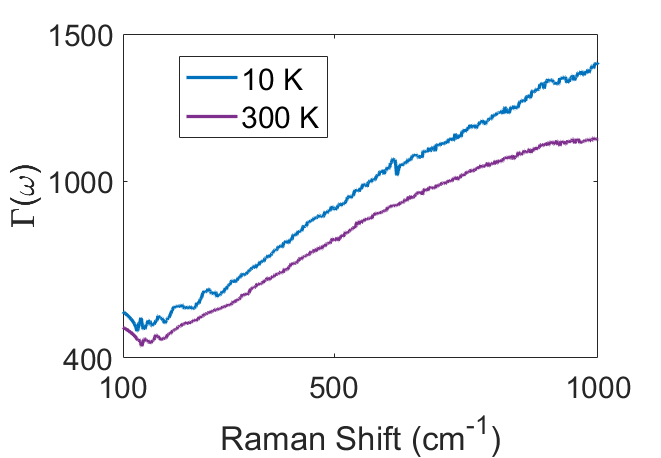}
    \includegraphics[width=0.65\linewidth]{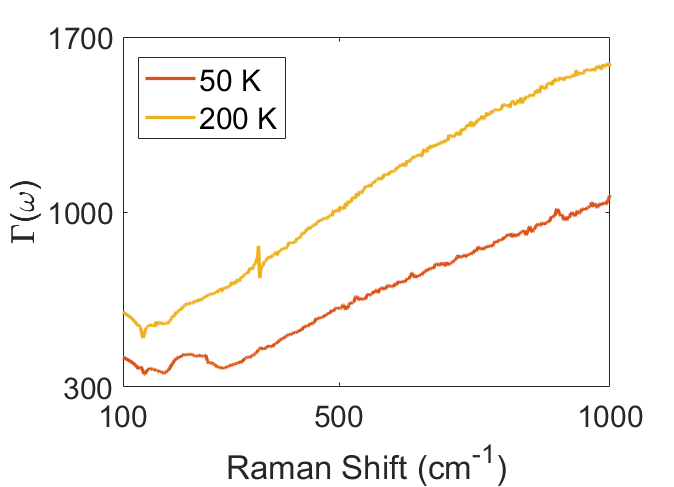}
    \includegraphics[width=0.65\linewidth]{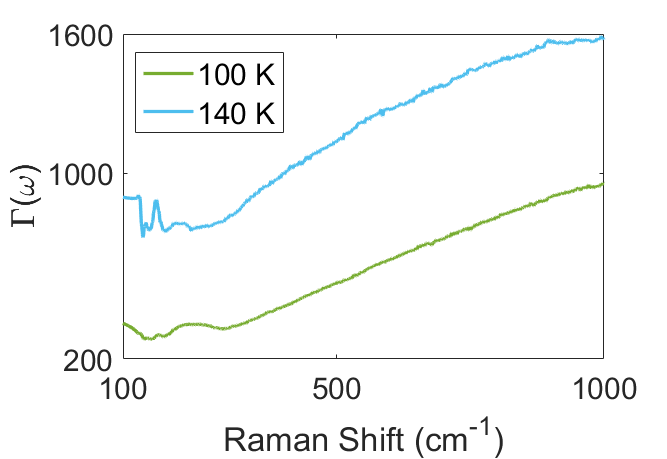}
    \caption{Plots of relaxation rate from memory function formalism. Each plot compares the relaxation rate as a function of frequency for one temperature value with and without the new peak ``N'' in the Raman spectra. The relaxation rate is linear for the temperature values with the new peak ``N'' in the Raman spectra which is a signature of plasmonic scattering confirming the plasmonic origin of the ``N'' peak.}
    \label{memory-func-res}
\end{figure}

With the above expressions we have calculated the Raman relaxation rate and mass-enhancement factor of PdTe$_2$ for various temperature values from the experimental Raman spectra. The results are presented in Fig. \ref{memory-func-res}. \\

\begin{figure}
    \centering
    \includegraphics[width=0.75\linewidth]{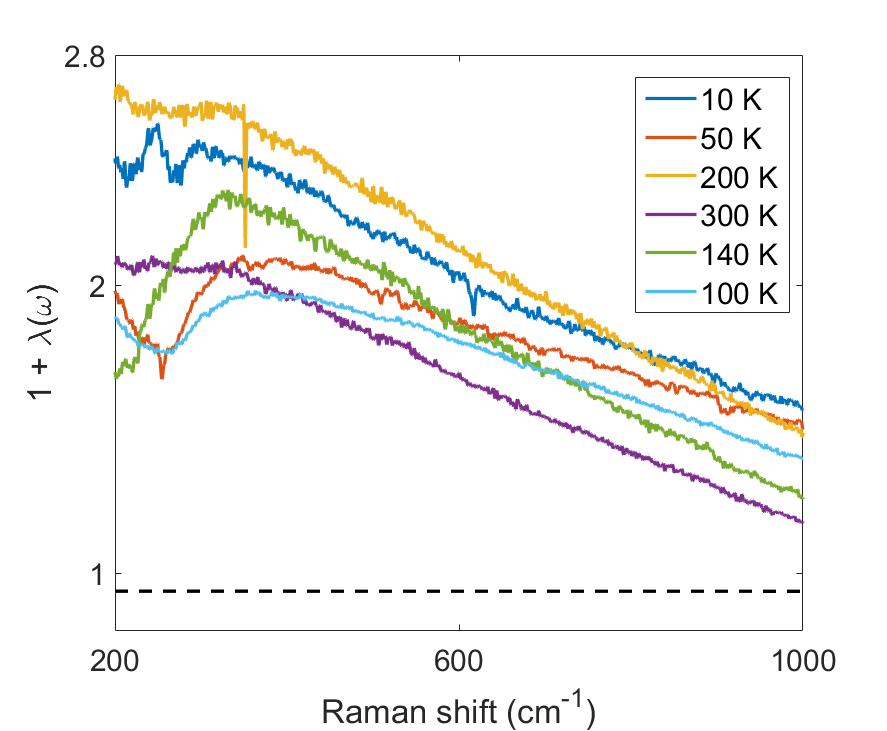}
    \caption{The mass enhancement factor as a function of frequency for different temperatures. For temperatures without the ``N'' peak in the Raman spectra, the mass enhancement factor drops off quickly then for the temperatures with the ``N'' peak present. }
    \label{memory-func-res-2}
\end{figure}

Before we discuss the results illustrated in Figs. \ref{memory-func-res} and \ref{memory-func-res-2}, we will briefly discuss certain benchmarks with which we validate the legitimacy of our numerical calculations. Most importantly the spectral dependence that we obtain for $\Gamma(\omega)$ and $\lambda(\omega)$ should have to be describable by sufficiently smooth functions and so is the case in our results. Also the results should not drastically depend on the value of the upper cut-off wave vector that is used for performing the numerical integrations. A reasonable cut-off prescribed is 3-5 times the upper limit of the region of interest. We have taken the cut-off to be 10 times the upper limit ($1000\,cm^{-1}$). Finally at higher frequencies, the value of the mass enhancement factor $(1 + \lambda(\omega))$, should asymptotically approach the value of 1, as $\lambda(\omega)$ reaches zero in a system of free non-local carriers. \\    

We will now discuss the results that we obtained through our phenomenological model. Based on the plots of Raman relaxation rate and mass-enhancement factor of PdTe$_2$ at different temperatures presented in Figs. \ref{memory-func-res} and \ref{memory-func-res-2}, we can infer the following:
\begin{itemize}
    \item The bifurcation in to lower and higher temperature made by us is based on the reference to the characteristic temperature ($\sim 100 \,K$), which corresponds to the  appearance of the third peak in the experimental Raman spectra data. Temperatures below and above the characteristic temperature are termed  as lower and higher temperatures respectively.
    \item The Raman relaxation rate ($\Gamma(\omega)$) for lower temperatures (10 K, 50 K, 100 K) is a linear function of the frequency ($\omega$) up to 1000 $cm^{-1}$. 
    \item The Raman relaxation rate ($\Gamma(\omega)$) for higher temperatures (140 K, 200 K, 300 K) is not a linear function of the frequency ($\omega$) and begins to saturate around 1000 $cm^{-1}$. This is indicative of the phonon scattering. In phonon scattering the relaxation rate saturates at the Debye wavenumber (for PdTe$_2$: 1408 cm$^{-1}$\cite{Hooda_2018}), which is evident from the relaxation rates for the above mentioned temperature range in Fig. \ref{memory-func-res}.
    \item The mass enhancement factor ($1 + \lambda(\omega)$) decreases with increase in frequency ($\omega$) for all values of temperature and asymptotically reaches the value of one. However the drop is faster for higher temperature values (140 K, 200 K, 300 K) than that of the lower ones (10 K, 50 K, 100 K).   
    \item The drastic change in the behavior of the Raman relaxation rate and the mass enhancement factor with respect to the temperature and its direct correlation with the appearance of the third peak in the Raman spectra is very interesting. Particularly there is no sign of the existence of the Drude scale ($\omega_{D}$) for $T\lesssim100K$.  
\end{itemize}

Comparing the results obtained in this section with the ones obtained in the previous sections, we could decisively conclude that the new peak corresponds to a plasmon-like mode. We will discuss this in detail in the next section.

\section{Discussion}\label{sec:dis}

Based on the experimental, computational and theoretical analysis of the Raman spectra of PdTe$_2$, we have come to the following conclusions:\\

The Raman spectra of PdTe$_2$, at low temperatures reveal the existence of a new peak (denoted as ``N'' in Fig.\ref{int_rs}) that appears below 100 K, along with the two already reported peaks ``E$_g$'' and ``A$_{1g}$''. The new peak is broader than the other two peaks and is centered around 250 $cm^{-1}$ wavenumber. The phonon frequency and linewidth of the two phonon peaks follow cubic anharmonicity below 100 K. Above 100 K the new peak fades and the the other two phonon modes exhibit anomalous behavior.\\

Our DFT calculations reveal that there are only two Phonon modes in PdTe$_2$, which are the already reported ``E$_g$'' and ``A$_{1g}$'' modes. This leads us to the conclusion that the third peak is not of phononic origin. It also cannot be a magnon mode since the system is non-magnetic. The only possibility left is that it originates from a plasmon-like scattering.\\

From the hint that the new mode may be emanating from the electronic Raman scattering\cite{cd3as2}, we construct a simple microscopic model of electrons coupled to a single plasmon mode. This is justified because we are informed from the experiments about the presence of only a single plasmon-like mode (below 100 K at 250 $cm^{-1}$ wavenumber). The model that we construct is similar to the Einstein model for electron-phonon coupling. Our calculations of the Raman relaxation rate from the memory function formalism for this simple model shows that in our region of interest (100-1000 $cm^{-1}$) the scattering rate depends linearly on the wavenumber.\\

We then calculate the Raman relaxation rate and mass enhancement factor with a phenomenological analysis of the experimental data. We observe that the Raman relaxation rate obtained from the phenomenological model also has a linear dependence on wavenumber for temperatures below 100 K. For temperatures above 100 K, the Raman relaxation rate saturates around the Debye wavenumber of 1400 $cm^{-1}$ for PdTe$_2$. This is expected as above 100 K, we have only phonon modes and scattering rate of phonon modes saturate at the Debye wavenumber. This clearly shows that below 100 K, in both the microscopic model and phenomenological analysis, we obtain a linear dependence of the Raman relaxation rate on the frequency. \\

The congruence of the results that we have obtained from the microscopic and phenomenological model allows us to conclude that the new peak appearing at temperatures below the characteristic temperature ($\sim 100 K$) is a plasmon-like mode.

\section{Conclusion}\label{sec5}

Our study highlights some exciting findings in PdTe$_2$, particularly the emergence of a new  mode centered around 250 $cm^{-1}$ wavenumber, below 100 K. Such a mode has been reported in the low temperature Raman spectra of similar material (Cd$_3$As$_2$) but the origin of this Raman mode has not been well understood. Our DFT calculations reveal that there are only two phonon modes in the system and the new third mode should be of different origin. Since the system is non-magnetic it cannot be a magnon mode. This leaves us with the possibility that it is a plasmon-like mode.\\

With this assumption,  we construct a simple microscopic model of electrons coupling to a single plasmon mode and obtain from the memory function formalism that the Raman relaxation rate should depend linearly on the frequency in our region of interest (100-1000 $cm^{-1}$). We then phenomenologically calculate the Raman relaxation rate from the experimentally obtained Raman spectra and find that the scattering rate thus obtained also shows a linear dependence on frequency. The agreement between the Raman relaxation rate calculated from the theoretical model and obtained phenomenologically from the experiments enables us to conclude that the new Raman mode in PdTe$_{2}$ is a plasmon-like mode.     


\vspace{1 cm}
\textbf{Acknowledgment}\\

\vspace{0.2 cm}
This work is supported by Physical Research Laboratory (PRL), Department of Space, Government of India. Computations were performed using the HPC resources (Param Vikram-1000 HPC) project at PRL. B.D. and J.P. acknowledge Dr. Som Narayan for the discussions regarding DFPT calculations. 


%
\bibliographystyle{unsrt}
\bibliography{bib}

\begin{thebibliography}{10}

\bibitem{Pdte2_int_1}
O.~J. Clark, M.~J. Neat, K.~Okawa, L.~Bawden, I.~Markovi\ifmmode~\acute{c}\else \'{c}\fi{}, F.~Mazzola, J.~Feng, V.~Sunko, J.~M. Riley, W.~Meevasana, J.~Fujii, I.~Vobornik, T.~K. Kim, M.~Hoesch, T.~Sasagawa, P.~Wahl, M.~S. Bahramy, and P.~D.~C. King.
\newblock Fermiology and superconductivity of topological surface states in {PdTe$_2$}.
\newblock {\em Phys. Rev. Lett.}, 120:156401, Apr 2018.

\bibitem{csy2}
Sonika, M.~K. Hooda, Shailja Sharma, and C.~S. Yadav.
\newblock Planar hall effect in {Cu} intercalated {PdTe$_2$}.
\newblock {\em Applied Physics Letters}, 119(26):261904, Dec 2021.

\bibitem{csy3}
Hancheng Yang, M.~K. Hooda, C.~S. Yadav, David Hrabovsky, Andrea Gauzzi, and Yannick Klein.
\newblock Anomalous charge transport of superconducting {Cu$_x$} {PdTe$_2$} under high pressure.
\newblock {\em Phys. Rev. B}, 103:235105, Jun 2021.

\bibitem{csy4}
Aastha Vasdev, Anshu Sirohi, M~K Hooda, C~S Yadav, and Goutam Sheet.
\newblock Enhanced, homogeneously type-{II} superconductivity in {Cu-intercalated PdTe$_2$}.
\newblock {\em Journal of Physics: Condensed Matter}, 32(12):125701, Dec 2019.

\bibitem{Pdte2_intro_1}
Han-Jin Noh, Jinwon Jeong, En-Jin Cho, Kyoo Kim, B.~I. Min, and Byeong-Gyu Park.
\newblock Experimental realization of type-{II} {Dirac} fermions in a {PdTe$_2$} superconductor.
\newblock {\em Phys. Rev. Lett.}, 119:016401, Jul 2017.

\bibitem{TMDC_intro_3}
H.~Leng, J.-C. Orain, A.~Amato, Y.~K. Huang, and A.~de~Visser.
\newblock Type-{I} superconductivity in the {Dirac} semimetal {PdTe$_2$} probed by $\ensuremath{\mu}\mathrm{SR}$.
\newblock {\em Phys. Rev. B}, 100:224501, Dec 2019.

\bibitem{TMDC_intro_10}
Yan Liu, Jian-Zhou Zhao, Li~Yu, Cheng-Tian Lin, Ai-Ji Liang, Cheng Hu, Ying Ding, Yu~Xu, Shao-Long He, Lin Zhao, Guo-Dong Liu, Xiao-Li Dong, Jun Zhang, Chuang-Tian Chen, Zu-Yan Xu, Hong-Ming Weng, Xi~Dai, Zhong Fang, and Xing-Jiang Zhou.
\newblock Identification of topological surface state in {PdTe$_2$} superconductor by angle-resolved photoemission spectroscopy.
\newblock {\em Chinese Physics Letters}, 32(6):067303, Jun 2015.

\bibitem{TMDC_intro_12}
Fucong Fei, Xiangyan Bo, Rui Wang, Bin Wu, Juan Jiang, Dongzhi Fu, Ming Gao, Hao Zheng, Yulin Chen, Xuefeng Wang, Haijun Bu, Fengqi Song, Xiangang Wan, Baigeng Wang, and Guanghou Wang.
\newblock Nontrivial {Berry} phase and {type-II Dirac} transport in the layered material $\mathrm{PdT}{\mathrm{e}}_{2}$.
\newblock {\em Phys. Rev. B}, 96:041201, Jul 2017.

\bibitem{RAMAN1928}
C.~V. Raman and K.~S. Krishnan.
\newblock The negative absorption of radiation.
\newblock {\em Nature}, 122(3062), Jul 1928.

\bibitem{Raman_intro_1}
Manish Jha, Amit~K. Bhojani, Sachin Pathak, Vishakha Kaushik, and Dheeraj~K. Singh.
\newblock Chapter 22 - {Raman spectroscopy: application to carbon-based nanomaterials}.
\newblock In {\em Applied Raman Spectroscopy}, pages 381--397. Elsevier, 2025.

\bibitem{Raman_intro}
R.~Merlin, A.~Pinczuk, and W.~H. Weber.
\newblock {\em Overview of Phonon Raman Scattering in Solids}, pages 1--29.
\newblock Springer Berlin Heidelberg, Berlin, Heidelberg, 2000.

\bibitem{Raman_intro_2}
Wei Ren, Helin Mei, Wenbo Sang, Mingshu Tan, Jianguo Si, Miao Liu, Jianting Ji, Feng Jin, Anmin Zhang, and Qingming Zhang.
\newblock Raman scattering study of electron-phonon coupling in superconducting {LaRu$_2$P$_2$}.
\newblock {\em Phys. Rev. B}, 110:144504, Oct 2024.

\bibitem{Raman_intro_3}
Zhang An-Min and Zhang Qing-Ming.
\newblock Electron—phonon coupling in cuprate and iron-based superconductors revealed by {Raman} scattering.
\newblock {\em Chinese Physics B}, 22(8):087103, Aug 2013.

\bibitem{plasmon}
Li-Lin Tay.
\newblock {\em Surface Plasmons}, pages 1481--1489.
\newblock Springer International Publishing, 2023.

\bibitem{plasmon2}
Xiang Wang, Sheng-Chao Huang, Shu Hu, Sen Yan, and Bin Ren.
\newblock Fundamental understanding and applications of plasmon-enhanced {Raman} spectroscopy.
\newblock {\em Nature Reviews Physics}, 2(5):253--271, May 2020.

\bibitem{Raman_intro_5}
N.~Kumar, N.V. Surovtsev, D.V. Ishchenko, P.A. Yunin, I.A. Milekhin, O.E. Tereshchenko, and A.G. Milekhin.
\newblock Resonance {Raman} scattering of topological insulators {Bi$_2$Te$_3$} and {Bi$_{2-x}$Sb$_x$Te$_{3-y}$Se$_y$ }thin films.
\newblock {\em Journal of Raman Spectroscopy}, 56(3), Mar 2025.

\bibitem{Raman_intro_4}
Xiangyu Hou, Xiao Tang, Yunjia Wei, Shanshan Wang, Qi~Hao, Jing-Min Hou, and Teng Qiu.
\newblock Role of dispersion relation effect in topological surface-enhanced {Raman} scattering.
\newblock {\em Cell Reports Physical Science}, 2(7):100488, Jul 2021.

\bibitem{lemmen3}
V.~Gnezdilov, Yu.~G. Pashkevich, H.~Berger, E.~Pomjakushina, K.~Conder, and P.~Lemmens.
\newblock Helical fluctuations in the {Raman} response of the topological insulator {Bi${}_{2}$Se${}_{3}$}.
\newblock {\em Phys. Rev. B}, 84:195118, Nov 2011.

\bibitem{Raman_intro_6}
J.~Knolle, Gia-Wei Chern, D.~L. Kovrizhin, R.~Moessner, and N.~B. Perkins.
\newblock Raman scattering signatures of {Kitaev} spin liquids in {A$_2$IrO$_3$} iridates with {A $=$ Na or Li}.
\newblock {\em Phys. Rev. Lett.}, 113:187201, Oct 2014.

\bibitem{Raman_intro_7}
Fang Liang, Hejun Xu, Xing Wu, Chaolun Wang, Chen Luo, and Jian Zhang.
\newblock Raman spectroscopy characterization of two-dimensional materials.
\newblock {\em Chinese Physics B}, 27(3):037802, Mar 2018.

\bibitem{Raman_intro_8}
Leandro~M. Malard, Lucas Lafeta, Renan~S. Cunha, Rafael Nadas, Andreij Gadelha, Luiz~Gustavo Cançado, and Ado Jorio.
\newblock Studying {2D} materials with advanced {Raman spectroscopy: CARS{,} SRS and TERS}.
\newblock {\em Phys. Chem. Chem. Phys.}, 23:23428--23444, Nov 2021.

\bibitem{lemmen2}
Dirk Wulferding, Peter Lemmens, Florian B\"uscher, David Schmeltzer, Claudia Felser, and Chandra Shekhar.
\newblock Effect of topology on quasiparticle interactions in the {Weyl} semimetal {WP$_{2}$}.
\newblock {\em Phys. Rev. B}, 102:075116, Aug 2020.

\bibitem{Taylor:51}
William~J. Taylor, A.~Lee Smith, and Herrick~L. Johnston.
\newblock A low temperature {Raman cryostat}.
\newblock {\em J. Opt. Soc. Am.}, 41(2):91--93, Feb 1951.

\bibitem{lemmen1}
Yu. Pashkevich, V.~Gnezdilov, P.~Lemmens, T.~Shevtsova, A.~Gusev, K.~Lamonova, D.~Wulferding, S.~Gnatchenko, E.~Pomjakushina, and K.~Conder.
\newblock Phase separation in iron chalcogenide superconductor {Rb$_{0.8+x}$Fe$_{1.6+y}$Se$_2$} as seen by {Raman} light scattering and band structure calculations.
\newblock {\em Low Temperature Physics}, 42(6), Jun 2016.

\bibitem{cd3as2}
A.~Sharafeev, V.~Gnezdilov, R.~Sankar, F.~C. Chou, and P.~Lemmens.
\newblock Optical phonon dynamics and electronic fluctuations in the {Dirac} semimetal $\mathrm{C}{\mathrm{d}}_{3}\mathrm{A}{\mathrm{s}}_{2}$.
\newblock {\em Phys. Rev. B}, 95:235148, Jun 2017.

\bibitem{Huang2016}
Xiaoting Huang, Yang Gao, Tianqi Yang, Wencai Ren, Hui-Ming Cheng, and Tianshu Lai.
\newblock Quantitative analysis of temperature dependence of {Raman} shift of monolayer {WS$_2$}.
\newblock {\em Scientific Reports}, 6(1):32236, Aug 2016.

\bibitem{csyadav1}
Sonika, Sunil Gangwar, Nikhlesh~Singh Mehta, Gargee Sharma, and C.~S. Yadav.
\newblock Chiral anomaly and positive longitudinal magnetoresistance in the {type-II Dirac} semimetals {A$_{x}$PdTe$_{2}$ (A $=$ Cu,Ag)}.
\newblock {\em Phys. Rev. B}, 108:245141, Dec 2023.

\bibitem{mmformalism_2}
W.~G\"otze and P.~W\"olfle.
\newblock Homogeneous dynamical conductivity of simple metals.
\newblock {\em Phys. Rev. B}, 6, Aug 1972.

\bibitem{mmformalim_1}
M.~Opel, R.~Nemetschek, C.~Hoffmann, R.~Philipp, P.~F. M\"uller, R.~Hackl, I.~T\"utt\ifmmode~\mbox{\H{o}}\else \H{o}\fi{}, A.~Erb, B.~Revaz, E.~Walker, H.~Berger, and L.~Forr\'o.
\newblock Carrier relaxation, pseudogap, and superconducting gap in high-${T}_{c}$ cuprates: A {Raman} scattering study.
\newblock {\em Phys. Rev. B}, 61, Apr 2000.

\bibitem{QE-2009}
Paolo Giannozzi, Stefano Baroni, Nicola Bonini, Matteo Calandra, Roberto Car, Carlo Cavazzoni, Davide Ceresoli, Guido~L Chiarotti, Matteo Cococcioni, Ismaila Dabo, Andrea {Dal Corso}, Stefano de~Gironcoli, Stefano Fabris, Guido Fratesi, Ralph Gebauer, Uwe Gerstmann, Christos Gougoussis, Anton Kokalj, Michele Lazzeri, Layla Martin-Samos, Nicola Marzari, Francesco Mauri, Riccardo Mazzarello, Stefano Paolini, Alfredo Pasquarello, Lorenzo Paulatto, Carlo Sbraccia, Sandro Scandolo, Gabriele Sclauzero, Ari~P Seitsonen, Alexander Smogunov, Paolo Umari, and Renata~M Wentzcovitch.
\newblock {QUANTUM ESPRESSO}: a modular and open-source software project for quantum simulations of materials.
\newblock {\em Journal of Physics: Condensed Matter}, 21(39):395502 (19pp), 2009.

\bibitem{sharma2020investigating}
Vaishali Sharma and Prafulla~K Jha.
\newblock Investigating the lattice dynamics and electronic structure of monolayer {PdTe$_2$}.
\newblock In {\em AIP Conference Proceedings}, volume 2265. AIP Publishing, 2020.

\bibitem{navindersir}
Navinder Singh.
\newblock {\em Electronic Transport Theories: From Weakly to Strongly Correlated Materials}.
\newblock CRC Press, 2016.

\bibitem{Hooda_2018}
M.~K. Hooda and C.~S. Yadav.
\newblock Electronic transport properties of intermediately coupled superconductors: {PdTe$_2$} and {Cu$_{0.04}$PdTe$_2$}.
\newblock {\em Europhysics Letters}, 121(1):17001, Mar 2018.

\end{thebibliography}
\end{document}